\documentclass[prl,twocolumn,letterpaper,byrevtex,floatfix,showpacs,%
preprintnumbers,superscriptaddress,amsfonts,amssymb,amsmath]{revtex4}

\usepackage{mathptmx}               
\usepackage{url}                
\renewcommand*{\vec}{\mathbf}           
\newcommand*{\J}{\vec{J}}           

\begin{document}

\title{Photon orbital angular momentum in a plasma vortex}

\author{J.\,T. Mendon\c{c}a}
 \email{titomend@ist.utl.pt}
 \affiliation{%
 IPFN and CFIF, Instituto Superior T\'{e}cnico,
 Av.~Rovisco Pais 1, 1049-001 Lisboa, Portugal}

\author{B. Thid\'e}
 \altaffiliation[Also at ]{LOIS Space Centre, V\"axj\"o University,
 SE-351\,95 V\"axj\"o, Sweden}
 \affiliation{Swedish Institute of Space Physics, P.\,O. Box 537,
 SE-751\,21 Uppsala,
 Sweden}

\author{J.\,E.\,S. Bergman}
 \affiliation{Swedish Institute of Space Physics, P.\,O. Box 537,
 SE-751\,21 Uppsala,
 Sweden}

\author{S.\,M. Mohammadi}
 \affiliation{Swedish Institute of Space Physics, P.\,O. Box 537,
 SE-751\,21 Uppsala,
 Sweden}

\author{B. Eliasson}
 \affiliation{%
 Department of Physics,
 Ume{\aa} University,
 SE-901\,87 Ume{\aa},
 Sweden}%

\author{W.\,A. Baan}
 \altaffiliation[Also at ]{LOIS Space Centre, V\"axj\"o University,
 SE-351\,95 V\"axj\"o, Sweden}
 \affiliation{%
 ASTRON,
 P.\,O. Box 2,
 NL-7991 PD Dwingeloo,
 The Netherlands and V\"axj\"o University, SE-351\,95 V\"axj\"o, Sweden}%

\author{H.\,Then}
 \affiliation{%
 Institute of Physics,
 Carl-von-Ossietzky Universit\"at Oldenburg,
 D-261\,11 Oldenburg,
 Germany}%

\date{\today}

\begin{abstract}

We study theoretically the exchange of angular momentum between a photon
beam and a plasma vortex, and demonstrate the possible excitation of
photon angular momentum states in a plasma.  This can be relevant to
laboratory and space plasma diagnostics; radio astronomy
self-calibration; and generating photon
angular momentum beams.  A static plasma perturbation with helical
structure, and a rotating plasma vortex are studied in detail and a
comparison between these two cases, and their relevance to the physical
nature of photon OAM, is established.

\end{abstract}

\pacs{52.35.We,52.70.Gw,42.50.Tx,41.20.Jb}

\maketitle


It is well known that photons can carry not only intrinsic spin angular
momentum (SAM), which is associated with their polarization state, but
also extrinsic orbital angular momentum (OAM) \cite{allen}.  The
existence of photon angular momentum has always been recognized on
theoretical grounds, and was first experimentally demonstrated in the
1930's \cite{beth36}.  While the experimental results led to quite some
discussions about their proper interpretation \cite{jauch}, they did not
excite much curiosity at the time as to their utilization.  It is also
known that quantum OAM states are associated with spherical wave
functions \cite{landau}, and can be excited by pointlike sources.  Only
recently photon OAM started receiving considerable attention, when it
was found that they can be associated not only with spherical waves
but also with cylindrical waves that can be easily produced by laser
sources.

The demonstration that in a laser beam the Laguerre-Gaussian modes
correspond to well defined OAM modes, and that these photon modes cannot
only be measured as a photon beam property \cite{harris,padgett}, but
can also be detected at the single photon level \cite{leach}, explains
the relevance of the present research on photon OAM states.  Utilization
of photon OAM states in the low frequency ($\lesssim1$~GHz) radio wave
domain, allowing digital control of the signals, was recently proposed
in Ref.~\onlinecite{thide}, as an additional method for characterizing
and studying radio sources.

The possibility of studying space plasma vorticity remotely by
measuring the OAM of radio beams interacting with the vortical
plasma was pointed out in Ref.~\onlinecite{thide2}.  Here we
analyze this possibility theoretically by studying the exchange of
angular momentum between a plasma medium and a photon beam.  Such
an analysis extends some of the more recent studies of optical
effects associated with photon OAM into the plasma physics domain,
and can be relevant for future plasma diagnostics, both in
laboratory and in space plasma.  A good understanding of the coupling
between plasma vorticity and radio beam OAM will pave the way for
improved self-calibration techniques in radio astronomy, and for finding
new methods of generating electromagnetic beams that carry OAM.

Two plasma situations will be analyzed in detail.  First, we
consider a photon beam propagating in a static plasma perturbation
with a helical vortex structure.  We will see that even in this
simple case, higher OAM states of the photon can be excited.
Secondly, we consider the interaction of the photon beam with a
rotating plasma vortex, characterized by a rotational angular
frequency $\Omega$. In contrast with the first case, the
excitation of higher order OAM states in the second situation is
accompanied by a frequency shift by multiples of $2\Omega$.


In the case of a static plasma vortex, we consider transverse
electromagnetic waves propagating in a isotropic plasma, where ion
motions will be neglected. In general terms, we can describe these
waves by the electric field propagation equation
\begin{equation}
\label{eq:2.1}
\Bigl(\nabla^2-\frac{1}{c^2}\frac{\partial^2}{\partial t^2}\Bigr)\vec{E}
 = \mu_0\frac{\partial\J}{\partial t}
\end{equation}
where the current $\J=-en\vec{v}$ is determined by the electron fluid equations
\begin{equation}
\label{eq:2.2}
\frac{\partial n}{\partial t} + \nabla\cdot n\vec{v} = 0
\; , \quad
\frac{\partial\vec{v}}{\partial t} + \vec{v}\cdot\nabla\vec{v}
 = -\frac{e}{m}(\vec{E} + \vec{v}\times\vec{B})
\end{equation}
Thermal and relativistic mass effects are ignored.  In the presence of a
static plasma vortex, we can define the mean electron density and
velocity by
\begin{equation}
\label{eq:2.3}
n =  n_0 + \tilde{n} (r, z) \cos (l_0 \varphi + q_0 z) \quad , \quad
\vec{v} = \vec{v}_0 (\vec{r},t) + \delta \vec{v}
\end{equation}
where $n_0$ and $\vec{v}_0$ describe the background plasma
conditions and $\delta \vec{v}$ is the perturbation associated
with the propagating electromagnetic wave.  We note that the
plasma helix vortex density perturbation, described in cylindrical
coordinates $\vec{r}\equiv(r,\varphi,z)$ depends on the distance
with respect to the vortex axis of symmetry and is allowed to vary
slowly along $z$, on a scale much longer than the spatial period
$z_0=2\pi/q_0$.  For a typical double vortex, we will have
$l_0=1$.  Plasma rotation is ignored for the moment, but will be
considered further below.  In the case of a static helical
perturbation, the plasma current is
$\vec{J}=-en_0(\vec{r})\delta\vec{v}$, and the propagation
equation can take the form
\begin{equation}
\label{eq:2.4b}
\Bigl\{\nabla^2 - \frac{1}{c^2}\frac{\partial^2}{\partial t^2}
 -\frac{\omega_{p0}^2}{c^2}\left[1+\epsilon(r,\varphi,z)\right]\Bigr\}\vec{E}=0
\end{equation}
where
\begin{equation}
\label{eq:2.4c}
\omega_{p0}^2 = \frac{e^2 n_0}{\epsilon_0 m} \;,\quad
 \epsilon (r,\varphi,z) = \frac{\tilde{n}(r,z)}{n_0}\cos(l_0\varphi+q_0z)
\end{equation}
We further assume wave propagation along the vortex axis $Oz$, and consider
solutions of the form
\begin{equation}
\label{eq:2.5}
\vec{E}(\vec{r},t) = \vec{A}(\vec{r})\exp
 \Bigl[-i \omega t + i\int^z k(z') dz' \Bigr]
\end{equation}
where $\omega$ is the wave frequency, and the wave amplitude $\vec{A}(\vec{r})$ varies only slowly along $z$ and satisfies
\begin{equation}
\label{eq:2.5b}
|\partial^2 \vec{A}/\partial z^2| \;\ll\; |2k\partial\vec{A}/\partial z|
\end{equation}
We can then reduce the wave equation (\ref{eq:2.4b}) to the perturbed
paraxial equation
\begin{equation}
\label{eq:2.6}
\Bigl[\nabla_\perp^2 + 2 i k \frac{\partial}{\partial z}
 - \frac{\omega_{p0}^2}{c^2}\epsilon(r,\varphi,z)\Bigr]\vec{A} = \vec{0}
\end{equation}
with
\begin{equation}
k^2 = \frac{1}{c^2} (\omega^2 - \omega_{p0}^2 ) \label{eq:2.6b}
\end{equation}
In the absence of the vortex perturbation, equation (\ref{eq:2.6})
would reduce to the usual paraxial optical equation.  In this
case, a general solution can be represented in a basis of
orthogonal Laguerre-Gaussian modes, according to the expansion
\begin{equation}
\vec{A} (r, \varphi, z) = \sum_{pl} A_{pl} (r, z) e^{i l \varphi} \exp
\Bigl( -\frac{r^2}{2 w^2} \Bigr) \vec{e}_{pl}
\label{eq:2.7} \end{equation}
where $w \equiv w (z)$ is the beam waist, $\vec{e}_{pl}$ are unit polarization vectors, and the amplitudes $A_{pl}$ are defined by
\begin{equation}
\label{eq:2.8}
A_{pl}(r,z) = A_{pl}(z)\frac{\sqrt{2}}{w}
 \Bigl[\frac{p!}{(l+p)!}\Bigr]^{1/2} x^{|l|/2} L_p^{|l|}(x)
\end{equation}
where $L_p^{|l|}(x)$ are the associated Laguerre polynomials, with
$x=r^2/w^2$.  The integers $p$ and $l$ represent the radial and the
azimuthal (quantum) numbers, respectively.  Using equation
(\ref{eq:2.7}) in equation (\ref{eq:2.5}), we can say that the total
electric field is represented by a superposition of Laguerre-Gaussian
states, in the form
\begin{equation}
\label{eq:2.9}
\vec{E} (\vec{r},t) = \sum_{pl}\vec{E}_{pl}(\vec{r})
 \exp\left(- i\omega t + i\int^z k (z') dz'\right)
\end{equation}
with
\begin{equation}
\label{eq:2.9b}
\vec{E}_{pl}(\vec{r}) = \vec{A}_{pl}(z)F_{pl}(r,\varphi)
\end{equation}
such that
\begin{equation}
\label{eq:2.10}
F_{pl}(r,\varphi)
 \propto\left(\frac{r^2}{w^2}\right)^{|l|/2}L_p^{|l|}
 \left( \frac{r^2}{w^2}\right)e^{il\varphi}\exp\left(-\frac{r^2}{2 w^2}\right)
\end{equation}
obeying the orthogonality condition
\begin{equation}
\label{eq:2.10b}
\int_0^\infty rdr\int_0^{2 \pi} d\varphi F_{pl}^*F_{p'l'}
 = \delta_{pp'}\delta_{ll'}
\end{equation}
In the general case, when a vortex perturbation
$\epsilon(r,\varphi,z)$ is present, these modes will be coupled to
each other, through the relation
\begin{equation}
\label{eq:2.11}
\frac{\partial}{\partial z}A_{pl}(z)
 = \frac{i}{2 k c^2}\sum_{p'l'}K(pl,p'l')A_{p'l'}
\end{equation}
where the coupling coefficients are defined by
\begin{equation}
\label{eq:2.12}
K(pl,p'l') = \omega_{p0}^2\int_0^\infty rdr
 \int_0^{2 \pi} d\varphi F_{pl}^*F_{p'l'}\epsilon (r,\varphi)
\end{equation}
In the simplest case where $\epsilon$ is only dependent on the
azimuthal angle $\varphi$, the coupling coefficients reduce to
\begin{equation}
\label{eq:2.13}
K (pl, p'l') = \omega_{p0}^2 \delta_{pp'}
 \int_0^{2 \pi}\epsilon(\varphi)e^{i(l'-l)\varphi} d\varphi
\end{equation}
This expression remains valid when the radial scale of the plasma
vortex is much larger than the photon beam waist $w (z)$.

As a special case, we consider a photon beam with no initial OAM,
which can be described by $\vec{E}_{pl}=0$ for $l\neq0$. Here the
mode coupling can be assumed to be sufficiently weak such that the
zero OAM mode is dominant over the entire interaction region, such
that $|E_{p0}|\gg|E_{p'l'\neq 0}|$.  For a helical static plasma
perturbation as defined by equation (\ref{eq:2.4b}), we then
obtain
\begin{equation}
\label{eq:2.14}
K(pl,p'l') = \pi\omega_{p0}^2 \frac{\tilde{n}}{n_0}\delta_{pp'}
 \left[\delta_{l',-l_0} e^{i q_0 z} + \delta_{l',l_0} e^{- i q_0 z}\right]
\end{equation}
Replacing this in the coupled mode equation (\ref{eq:2.11}), and
integrating over the axial coordinate $z$, we obtain, for
$A_{pl}(0)=A(0)\delta_{l0}$, and assuming the same polarization state
for all the interacting modes, the following expression for the field
mode amplitudes
\begin{equation}
\label{eq:2.15}
A_{p,\pm l_0} (z) = i \frac{\pi A (0)}{2 c^2}
 \int_0^z\frac{\omega_{p0}^2(z')}{k(z')}
 \frac{\tilde{n}(z')}{n_0}e^{\mp iq_0 z'} dz'
\end{equation}
This expression describes the rate of transfer of OAM from the
static plasma vortex to the electromagnetic field.  We notice that
such a transfer is inhibited after a distance of order $2 \pi /
q_0$, the helical path length.  The favorable case is therefore
that of an interaction distance shorter than this length, or in
the limit, a plasma structure with no axial periodicity.  This
picture will qualitatively change for the case of a time dependent
plasma perturbation considered below.

Equation (\ref{eq:2.15}) is only valid when the transfer of OAM is
small, such that the amplitude of the initial Gaussian mode
$A_{p0}$ can be considered constant along the axis. In order to
derive a more general solution where the amplitude of the
initially excited mode is allowed to change, we can go back to the
coupled mode equations (\ref{eq:2.11}) and (\ref{eq:2.12}) and
assume a nearly constant coupling coefficient. Writing these
equations in a simplified but obvious new notation where we drop
the radial index $p$, we obtain
\begin{equation}
\label{eq:2.16}
\frac{\partial}{\partial z} A_l
 =i K_0 \left[ A_{l-l_0} e^{i q_0 z} + A_{l+l_0} e^{- i q_0 z} \right]
\end{equation}
with
\begin{equation}
\label{eq:2.16b}
K_0 = \frac{\pi\omega_{p0}^2}{2 k c^2}\frac{\tilde{n}}{n_0}
\end{equation}
assumed constant along the interaction length.  For $q_0\rightarrow0$,
this can be solved in terms of Bessel functions
\begin{equation}
A_{l_i+\nu l_0} (z) = i^{- \nu} A (0) J_\nu ( 2 K_0 z)
\label{eq:2.17} \end{equation} where $l_i$ is the initial orbital
angular momentum state of the electromagnetic beam, and $\nu$ is
an integer.  This solution clearly shows the decay of the initial
state $l_i$ over all the other states ($l_i + \nu l_0$), on a
length scale approximately determined by the inverse of the
coupling constant $K_0$.


Secondly we consider the case of rotating plasma vortex, where the
plasma density profile is assumed to be homogeneous and constant,
$\epsilon(r,\varphi) = 0$, but the plasma is allowed to rotate
with respect to the laboratory frame (the frame attached to the
photon beam source), with an angular velocity $\Omega$, around the
axis $Oz$.  This means that plasma particles located at the
transverse coordinates $(r,\varphi)$ have an unperturbed velocity
given by
\begin{equation}
\label{eq:3.1}
\vec{v}_0 = v_0 (r) \vec{e}_\theta = v_0 (r) [ - \sin (\Omega t +
\varphi) \vec{e}_x + \cos (\Omega t + \varphi) \vec{e}_y ]
\end{equation}
In order to determine the perturbed velocity associated with the wave
field, $\delta \vec{v}$, we linearize the electron momentum equation with respect to this quantity, and write
\begin{equation}
\label{eq:3.2}
 \left(\frac{\partial}{\partial t}+\vec{v}_0\cdot\nabla\right)\delta \vec{v}
 +\delta\vec{v}\cdot\nabla\vec{v}_0
 = -\frac{e}{m}\left(\vec{E}+\vec{v}_0\times\vec{B}\right)
\end{equation}
For very slow rotation velocities, we can neglect the magnetic field
term.  This term would lead to a current contribution, oscillating at
frequencies $\omega \pm \Omega$, which is ignored here.  The remaining
equation shows that the plasma rotation introduces a frequency shifted
term, oscillating at plus or minus twice the rotation frequency
$\Omega$.  As a consequence, the resulting plasma current has three
different frequency components, and can be written as
\begin{equation}
\vec{J} (t)  = - e n_0 \vec{v} (t) = \sum_{\nu = 0, \pm1} \vec{J}_\nu \exp [ - i (\omega + 2 \nu \Omega ) t  + 2 i \nu \varphi ]
\label{eq:3.6} \end{equation}
which introduces a coupling between three different frequency modes
$\omega$ and $(\omega \pm 2 \Omega)$.  For a circularly polarized
electric field $\vec{E} = E_\pm \vec{e}_\pm$, we can write
\begin{equation}
\vec{J}_0 = - \frac{e^2 n_0 E_\pm}{\omega m} \left( 1 \pm \frac{i}{2
\omega} \frac{\partial v_0}{\partial r}  \right) \vec{e}_\pm
\label{eq:3.7} \end{equation}
and
\begin{equation}
\vec{J}_{\pm 1} = \pm i \frac{e^2 n_0}{2 \sqrt{2} \omega m}
\frac{E_\mp}{(\omega \pm \Omega)} \frac{\partial v_0}{\partial r} \vec{e}_\pm
\label{eq:3.7b} \end{equation}
Here we have defined
$\vec{e}_\pm = (\vec{e}_x \pm i \vec{e}_y) / \sqrt{2}$. For arbitrary
polarization, we use $\vec{E} = E_+ \vec{e}_+ + E_- \vec{e}_-$.
Notice that the polarization states $\vec{e}_\pm$ correspond to the two
photon spin states.  The above expressions show that a rotating plasma
responds differently to the different spin states.  More interestingly,
these spin states become coupled by the frequency shifted terms
determined by $\vec{J}_\pm$.  Here however, we concentrate our attention
on the photon OAM states, and consider a linearly polarized field, with
equal contributions from the two spin states.

These expressions for the electron current suggest the use of the
following wave solution
\begin{equation}
\vec{E} (\vec{r}, t) =  \sum_{npl} \vec{E}_{npl} (\vec{r}, t) \exp \left[- i (\omega + n \Omega) t + i \int^z k (z') d z' \right]
\label{eq:3.8} \end{equation}
with
\begin{equation}
\vec{E}_{mpl} (\vec{r}, t) =  \vec{A}_{npl} (z) F_{pl} (r, \varphi)
\label{eq:3.8b} \end{equation}
where $F_{np}$ are determined by equation (\ref{eq:2.10}).  This is a
straightforward generalization of the solutions (\ref{eq:2.9}) and
(\ref{eq:2.9b}).  Replacing this form of solution in the modified
paraxial equation, we obtain new coupled mode equations for the slowly
varying field amplitudes
\begin{equation}
\frac{\partial}{\partial z} A_{npl} = \frac{i}{2 k_n c^2} \sum_\nu \sum_{n'p'l'} K_\nu (npl, n'p'l') A_{n'p'l'} e^{i  (k_{n'} - k_n) z}
\label{eq:3.9} \end{equation}
where
\begin{equation}
k_n = \frac{1}{c} (\omega_n^2  - \omega_{p0}^2 )^{1/2}
\label{eq:3.9b} \end{equation}
The coupling coefficients are now defined as
\begin{multline}
K_\nu (npl, n'p'l') \\
 = - i\omega_{n - \nu}\int_0^\infty r dr
 \int_0^{2 \pi} d\varphi\sigma_\nu(r) F_{npl}^*F_{n'p'l'}e^{2 i \nu \varphi}
\label{eq:3.10} \end{multline}
where the quantities $\sigma_\nu (r)$ possibly depend on the radial
coordinate through the rotation frequency $\Omega (r)$.  We assume here
a constant rotation frequency (over the beam waist distance) and also
assume linear polarization of the incident wave beam in the $Ox$
direction, which leads to
\begin{equation}
\sigma_0 = - \frac{\omega_{p0}^2}{\omega}
\quad , \quad
\sigma_{\pm 1} = i  \frac{\omega_{p0}^2}{(\omega \pm \Omega)} \frac{\partial v_0}{\partial r}
\label{eq:3.11} \end{equation}
We are then left with
\begin{equation}
K_\nu (npl, n'p'l') = - 2 \pi i \omega_{n' - \nu}  \sigma_\nu \delta_{pp'} \delta_{l,l'+2 \nu}
\label{eq:3.11b} \end{equation}
Replacing this in the coupled mode equations, we get, for a well defined
but unspecified quantum number $p$, the following result
\begin{equation}
\frac{\partial}{\partial z} A_{npl} = \sum_{\nu = \pm 1}  \bar{K}_{n \nu} A_{n, l + 2 \nu} e^{i \Delta_\nu z}
\label{eq:3.12} \end{equation}
with $\Delta_\nu = (k_{n - \nu} - k_n)$, and
\begin{equation}
\label{eq:3.12b}
\bar{K}_{n \nu} = \pi \frac{\omega_{n - \nu}}{k_n c^2} \sigma_\nu
\end{equation}
This can be simplified further, by noting that the vortex rotation
frequency is much lower than the incident wave frequency
$\Omega\ll\omega$, which allows us to replace, for moderate values of
the integer $n$, the coupling coefficients $\bar{K}_{n\nu}$ by a
constant value, $\bar{K}=(\pi\omega_{p0}^2 / c \omega^2)
(\partial v_0/\partial r)$, leading to
\begin{equation}
\label{eq:3.13}
\frac{\partial}{\partial z} A (n, l)
 = \bar{K} \bigl[A(n-2,l+2)e^{-i\Delta z}
  + A (n + 2, l - 2) e^{i \Delta z} \bigr]
\end{equation}
with $\Delta \simeq 2 \Omega / c$.  In contrast to the case of
static perturbations (which only couples states with equal
frequency but different orbital angular momentum), the case of a
rotating but uniform plasma couples both frequency and angular
momentum states. According to equation (\ref{eq:3.13}), a given
mode $(\omega_n, l)$ can decay into the modes $(\omega_{n+2},
l-2)$ and $(\omega_{n-2}, l+2)$.

For initial photon states with frequency $\omega = \omega_0$ and no
orbital angular momentum $(l_0 = 0)$, we obtain from equation
(\ref{eq:3.13}) the following cascading decay equations
\begin{equation}
\frac{\partial}{\partial z} A_\alpha =  \bar{K} \left[ A_{\alpha+1} e^{-i \Delta z} + A_{\alpha-1} e^{i \Delta z} \right]
\end{equation}
with $A_\alpha \equiv A (n = 2 \alpha, l = - 2 \alpha)$.  Again, in the
limit of a negligible dephasing, $\Delta \rightarrow 0$, we can get
simple Bessel function solutions, determined by
\begin{equation}
A (n_0 + 2 n, l_0 - 2 n) = (-1)^n A (0) J_n ( 2 \bar{K} z)
\label{eq:3.14} \end{equation}
where $A (0) \equiv A (n_0, l_0)$ is the amplitude of the initial photon
state.  Notice the similarity with our previous solutions for a static
vortex.

Finally, we could consider plasma vortex perturbations with some
parallel wavelength, of the form $\exp (i q_0 z)$.  In this case, we
would have to replace $\Delta$ by $(\Delta - q_0) z$ in the previous
mode coupled equations.  This means that a perfect phase matching
between the various photon modes can be achieved (at least in the limit
$\Omega \ll \omega_n$), for $\Delta = q_0$.


In this work we have considered photon orbital angular momentum,
for electromagnetic waves propagating in a plasma.  To our
knowledge, this is the first theoretical work dealing with photon
OAM in the context of plasma vorticity.  We have studied the
propagation of electromagnetic wave beams in an isotropic plasma,
and we have discussed two distinct physical conditions:  a helical
disturbance in static plasma, and a rotating plasma vortex.  We
have shown that, in both cases, a cascading process of OAM
transfer between the plasma and a photon beam can be achieved. For
static plasma perturbations with a finite helicity there is no
photon frequency shift.  Only the total angular momentum of the
photon beam is modified.  In contrast, for a rotating plasma
perturbation without any helical perturbation, photon OAM states
are excited with photon frequency shifts that are multiples of
twice the plasma rotation frequency.  These distinct features can
be used as an additional diagnostic method of the plasma
properties, and can be useful in the context of both space and
laboratory plasma.


\end{document}